\documentclass[preprint,showpacs,preprintnumbers,amsmath,amssymb]{revtex4}

\usepackage{graphicx}
\usepackage{bm} 
\usepackage[]{amsfonts, graphicx, amsmath}
\DeclareGraphicsExtensions{.eps,.ps,.eps.gz,.ps.gz,.eps.Z}

\newcommand{\ra}{\rangle}
\newcommand{\la}{\langle}
\newcommand{\tr}{{\rm Tr}}

\begin{document}

\title{Quantum walks with infinite hitting times}
\author{\sc Hari Krovi}\email{krovi@usc.edu}
\author{Todd A. Brun}\email{tbrun@usc.edu}
\affiliation{Communication Sciences Institute, University of Southern California, \\
Los Angeles, California 90089, USA}
\date{\today}

\begin{abstract}
Hitting times are the average time it takes a walk to reach a given final vertex from a given starting vertex. The hitting time for a classical random walk on a connected graph will always be finite. We show that, by contrast, quantum walks can have infinite hitting times for some initial states. We seek criteria to determine if a given walk on a graph will have infinite hitting times, and find a sufficient condition, which for discrete time quantum walks is that the degeneracy of the evolution operator be greater than the degree of the graph.  The set of initial states which give an infinite hitting time form a subspace.  The phenomenon of infinite hitting times is in general a consequence of the symmetry of the graph and its automorphism group. Using the irreducible representations of the automorphism group, we derive conditions such that quantum walks defined on this graph must have infinite hitting times for some initial states. In the case of the discrete walk, if this condition is satisfied the walk will have infinite hitting times for any choice of a coin operator, and we give a class of graphs with infinite hitting times for any choice of coin.  Hitting times are not very well-defined for continuous time quantum walks, but we show that the idea of infinite hitting-time walks naturally extends to the continuous time case as well.

\end{abstract}

\pacs{03.67.Lx 05.40.Fb}

\maketitle

\section{Introduction}

Quantum walks have been defined and analyzed in the hope that they will lead to new quantum algorithms, as classical random walks have for classical algorithms. Classically, many probabilistic algorithms have the structure of a random walk; for instance, the most efficient algorithm for 3-SAT is based on a random walk \cite{Mot}. The search for new types of quantum algorithms is spurred by the fact that the use of quantum Fourier transform as an algorithmic tool---used in Shor's algorithm and related algorithms---seems to be restricted to Abelian groups and certain classes of non-Abelian groups for the hidden subgroup problem \cite{Shor}. Quantum walks may provide a new approach. So far they have been applied to the element distinctness problem \cite{ambainis}, and an alternative search algorithm \cite{shenvi} that has been shown to have a quadratic speed-up (the same as Grover's algorithm). In \cite{childs2}, it was shown that the quantum walk on the so-called ``glued trees'' graph reaches the final vertex from the initial vertex exponentially faster that a similar classical walk.

There are two kinds of quantum walks:  discrete time and continuous time. They are defined differently, and unlike the classical case, there is no natural limit which leads from one to the other. Discrete time quantum walks on regular graphs require an extra system, called the ``coin,'' whose Hilbert space dimension equals the degree of the graph. Quantum walks have been analyzed extensively for the line, the continuous time case in Refs.~\cite{childs2,farhi,childs} and the discrete time case in \cite{nayakV,bachWatrous,todd1,todd2}. Quantum walks on the hypercube have been studied in \cite{shenvi,mooreRussell,kempe,KroviBrun}.  Quantum walks have been defined for general (irregular) undirected graphs in \cite{ambainisReview} and \cite{viv}, and for directed graphs in \cite{montanaro}. Certain properties of quantum walks,  such as mixing time and sampling time, were defined by analogy to classical random walks and discussed in \cite{aharanov}. Hitting times were defined in \cite{kempe}, where an upper bound was found for the walk on the hypercube. A different definition of hitting time was defined in \cite{KroviBrun}, where the usual definition of a quantum walk as a unitary evolution has been replaced by a measured walk. In this walk, a measurement is performed at every step after the application of the unitary, to see if the particle has reached the final vertex or not. Recent reviews of quantum walks can be found in \cite{ambainisReview} and \cite{kempeReview}.  In this paper, we will primarily concentrate on discrete time walks, but will show that our results extend naturally to continuous time walks as well.

It was shown in \cite{KroviBrun} that for a certain initial state, the hitting time for a discrete quantum walk on the hypercube using the so called ``Grover'' coin has a hitting time that is exponentially faster than the classical hitting time on the same graph. By contrast, for the same initial condition with another coin---the Discrete Fourier Transform (DFT) coin---the hitting time was infinite, something that never occurs in a classical walk on a connected graph.  In this paper, we study this phenomenon of infinite hitting times in quantum walks, and show that it is generally caused by degeneracy of the evolution operator, and hence is linked to the symmetry.  Symmetries in the automorphism group of the graph lead to symmetries in the evolution operator, which in turn induce degeneracies in its eigenvalues. If the degeneracy is sufficiently great, some of these degenerate eigenvectors will span a subspace that has no overlap with the final vertex.  The projector onto this subspace commutes with both the evolution operator and the measurement. So, a walk whose initial state is in this subspace will never leave the subspace and will continue to have zero overlap with the final state for all times. A measurement performed at any time will never find the particle in the final state, leading to an infinite hitting time. Thus, infinite hitting times are a natural phenomenon for graphs that have certain kinds of symmetries.

We use the theory of irreducible representations to estimate the amount of degeneracy that a given group of symmetries produces. The use of representation theory to explain aspects of quantum walks on certain classes of graphs was also done in \cite{GerhardtWatrous}, where the behavior of mixing times of Cayley graphs on the symmetric group is explained based on its irreducible representations.  We will also use Cayley graphs to produce examples of quantum walks with infinite hitting times.

This paper is organized as follows. In Sec. II, we discuss quantum walks---both discrete time and continuous time---and then define Cayley graphs and their automorphism groups. In Sec. III, we review hitting times for discrete time quantum walks, and give examples where they can become infinite.  We discuss what properties of a quantum walk lead to infinite hitting times, and derive a sufficient condition based on the degeneracy of the evolution operator. In Sec. IV, we briefly review group representation theory, and use it to show that there exist graphs which have infinite hitting times for both discrete and continuous walks.  We provide examples in each case. In Sec. V, we discuss the symmetry of the graph and its influence on the quantum walk.

\section{Random walks on graphs}\label{walks}

\subsection{Quantum walks - discrete and continuous}

Quantum walks are primarily of two types. Depending on the way the evolution operator is defined, they can be either discrete time or continuous time. These two definitions of quantum walks are not exactly equivalent to the two types of classical random walks. While they are both based on the classical definitions, unlike the classical case the discrete quantum walk does not reduce to the continuous walk when we let the time step between repeated applications of the unitary tend to zero. This is because discrete time walks need an extra Hilbert space, called the ``coin''  space (from the idea that one flips a coin at each step to determine which way to walk), and taking the limit where the time step goes to zero does not eliminate this Hilbert space.  (But see \cite{Strauch06} for a different treatment of this limit.)  Therefore, the properties of discrete and continuous walks are different. Though there is no obvious reason why one should be preferred, in some cases it has been shown that coins make these walks faster \cite{AmbainisKempe}.

\subsubsection{Discrete time walks}

A discrete time quantum walk can broadly be defined as the repeated application of a unitary evolution operator on a Hilbert space whose size depends on the graph. This Hilbert space usually consists of the space of possible positions (i.e., the vertices) together with the space of possible directions in which the particle can move along from each vertex (the coin space). In this paper, we only consider $d$-regular, undirected and $d$-colorable graphs.  We will briefly review the definition of these graph properties.

A {\it regular} graph is one where every vertex is connected to the same number $d$ of other vertices.  This number is called the the {\it degree} of the graph.  A graph is {\it undirected} if for every edge between vertices A and B going from A to B, an edge goes from B to A as well.  In this case, we identify the edge from A to B with the edge from B to A, and consider them a single edge.  A regular, undirected graph with $N$ vertices of degree $d$ is considered $d$-{\it colorable} if the edges incident on every vertex can be numbered $1$ through $d$ such that every edge between two vertices has the same number at either end. Not all $d$-regular and undirected graphs can be $d$-colored. A simple example is the triangle graph where $N=3$ and $d=2$.  (See Fig.~1.)

\begin{figure}[t]
\begin{center}
\includegraphics[scale=0.8]{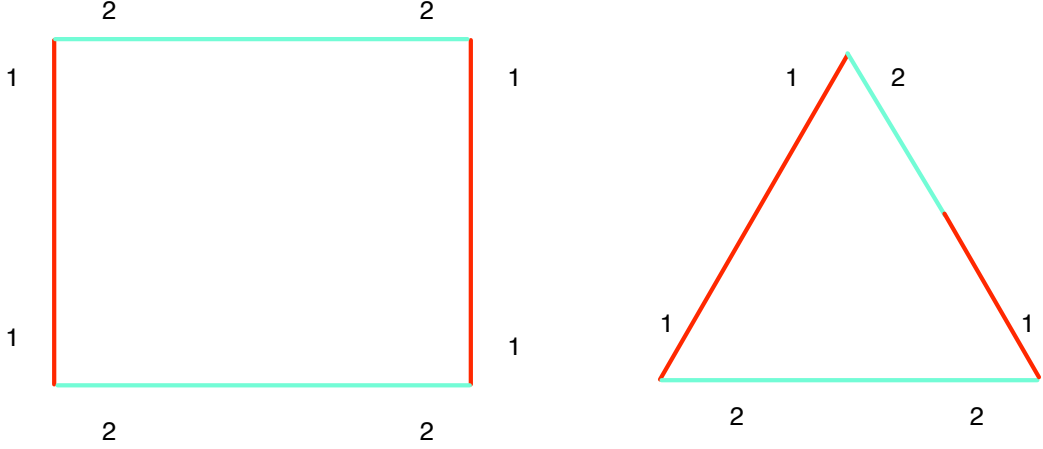}
\end{center}
\caption{(Color online) Examples of 2-colorable (square) and non-2-colorable (triangle) regular graphs of degree 2, where the colors are numbered 1 and 2.} \label{fig1}
\end{figure}

For $d$-regular, undirected and $d$-colored graphs, the Hilbert space of the walk is $\mathcal{H}^p\otimes\mathcal{H}^c$, i.e., the tensor product of the position and direction (or coin) space. The evolution operator $\hat{U}$ is given by $\hat{U}=\hat{S}(\hat{I}\otimes\hat{C})$, where $\hat{S}$ is called the shift matrix and $\hat{C}$ is the coin matrix. The shift matrix encodes the structure of the graph and is very similar to its adjacency matrix. The vertices, numbered $|0\ra$ through $|N-1\ra$, are basis states for the vertex Hilbert space $\mathcal{H}^p$ and the set of all directions from each vertex, numbered $|1\ra$ through $|d\ra$, are basis states for the coin Hilbert space $\mathcal{H}^c$. In this basis, the shift matrix for the graph can be given the explicit form:
\[
\hat{S} = \sum_v \sum_i |v(i),i\ra\la v,i| ,
\]
where $v(i)$ is the vertex connected to $v$ along the edge numbered $i$.

The coin matrix $\hat{C}$ acts only on the coin space, and ``flips'' the directions before the shift matrix is applied. Then $\hat{S}$ moves the particle from its present vertex to the vertex connected to it along the edge indicated by the coin direction. Though $\hat{C}$ can be any unitary matrix, usually coins with some structure are considered. The coins that we used in our previous analysis are the Grover coin $\hat{G}$ and the Discrete Fourier Transform (DFT) coin $\hat{D}$. The matrices for these coins are given by:
\begin{equation}
\hat{G} = 2|\Psi\ra\la\Psi|-I = \begin{pmatrix}
  \frac{2}{d}-1 & \frac{2}{d} & \ldots & \frac{2}{d} \\
  \frac{2}{d} & \frac{2}{d}-1 & \ldots & \frac{2}{d} \\
  \vdots & \vdots & \ddots & \vdots \\
  \frac{2}{d} & \frac{2}{d} & \ldots & \frac{2}{d}-1
\end{pmatrix} ,
\label{Grover_matrix}
\end{equation}
and
\begin{equation}
\hat{D}=\frac{1}{\surd{d}}\begin{pmatrix}
  1 & 1 & 1 & \ldots & 1 \\
  1 & \omega & \omega^2 & \ldots & \omega^{d-1} \\
  \vdots & \vdots & \vdots & \ddots & \vdots \\
  1 & \omega^{d-1} & \omega^{2(d-1)} & \ldots &
  \omega^{(d-1)(d-1)} ,
\end{pmatrix} ,
\end{equation}
where $|\Psi\ra=\frac{1}{\surd{d}}\sum_i|i\ra$ and $\omega=\exp(2\pi i/d)$.

\subsubsection{Continuous time walks}

Continuous time quantum walks were defined by Farhi et al. in \cite{farhi}. For an undirected graph $G(V,E)$, the unitary evolution operator is defined as $\hat{U}=\exp(i\hat{H}t)$, where $\hat{H}$ is taken to be the adjacency matrix of the graph. Here again, the vertices of the graph form a basis for the Hilbert space on which $\hat{U}$ is defined. This gives rise to the following Schr\"{o}dinger equation:
\begin{equation}
i\frac{d}{dt} \la v|\psi (t)\ra=\la v|\hat{H}|\psi (t) \ra.
\end{equation}
$\hat{H}$ is a symmetric matrix (and hence $\hat{U}$ unitary) if the graph is undirected. This walk has a structure very similar to that of continuous time Markov chains. The adjacency matrix $\hat{H}$, which acts as the Hamiltonian, is of the form:
\begin{equation}
H_{i,j}=\biggl\{ \begin{array}{cc} 1 & {\rm if}\ i\ {\rm and}\ j\ {\rm share\ an\ edge,} \\
0 & {\rm otherwise.} \end{array} .
\end{equation}
As can be seen, this walk has no coin and so the Hilbert space on which $\hat{U}$ acts is only the vertex space $\mathcal{H}^p$.

\subsection{Cayley graphs and automorphism groups}

Cayley graphs are defined in terms of a group $G$ and a set $S$ consisting of elements from $G$ such that the identity element $e\notin S$. Given $G$ and $S$, the resulting (right)-Cayley graph $\Gamma(G,S)$ is one whose vertices are labeled by the group elements, i.e., there is one vertex for every group element, and two vertices $g$ and $h$ are connected by a directed edge from $g$ to $h$ if $g^{-1}h\in S$, (see \cite{GrossTucker}). Another way to look at this definition is that from any vertex $g$ of a Cayley graph, there are $|S|$ outgoing edges, one to each of the vertices $gs$, $\forall s\in S$. A Cayley graph will be connected if and only if the set $S$ is a generating set for $G$, and it will be undirected if $s^{-1}\in S$, $\forall s\in S$. The degree of such a graph is $|S|$, the cardinality of the generating set. Finally, a $d$-regular Cayley graph can be $d$-colored if $s^2=1$, $\forall s\in S$, i.e., $s^{-1}=s$.

Examples of Cayley graphs on which quantum walks have been studied include the line $\Gamma(\mathcal{Z},\{1,-1\})$; the cycle $\Gamma(\mathcal{Z}_n,\{1,-1\})$; the hypercube $\Gamma(\mathcal{Z}_2^n,X)$ where the set $X$ is the set of canonical generators $\{(1,0,0,\cdots,0),(0,1,0,\cdots,0),\dots,(0,0,0,\cdots,1)\}$; and the graph on the symmetric group $\Gamma (S_n,Y)$, where $Y$ is a generating set for $S_n$.  Let us look at the hypercube as an example of a Cayley graph where quantum walks have been extensively studied.

Consider the hypercube, which has $|\mathcal{Z}_2^n|=2^n$ vertices each with a degree of $|X|=n$. The vertices can be labeled by an $n$-bit string from $(0,0,\cdots,0)$ through $(1,1,\cdots,1)$. Two vertices are adjacent if they differ only by a single bit. Vertex $\vec{v}$ is connected to $n$ vertices given by $\vec{v}\oplus\vec{s}$, $\forall \vec{s}\in X$, where $\oplus$ stands for the bit-wise XOR of the bit strings $\vec{v}$ and $\vec{s}$. One important property of the hypercube is that it can be $n$-colored, since $\vec{s}\oplus\vec{s} = \vec{e}$, $\forall \vec{s}\in X$ where $\vec{e} = (0,0,\cdots,0)$ is the identity element. The unitary evolution operator for a discrete walk on the hypercube becomes $\hat{U}=\hat{S}(\hat{I}\otimes \hat{C})$, where $\hat{S}$ has the form
\[
\hat{S}=\sum_{\vec{s}}\sum_{\vec{v}}|\vec{v}\oplus\vec{s}\ra\la \vec{v}|\otimes |\vec{s}\ra\la\vec{s}|.
\]
Since the vertices of the hypercube are bit strings, and adjacent vertices are those that differ by one bit, the shift matrix of the discrete walk on the hypercube has a natural form given by
\begin{eqnarray}
\hat{S} &=& \hat{X}\otimes\hat{I}\otimes\dots\otimes\hat{I}\otimes |\vec{s_1}\ra\la\vec{s_1}|
+ \hat{I}\otimes\hat{X}\otimes\dots\otimes\hat{I}\otimes\hat{|}\vec{s_2}\ra\la \vec{s_2}| \nonumber\\
&& + \ldots + \hat{I}\otimes\hat{I}\otimes\dots\otimes\hat{X}\otimes |\vec{s_n}\ra\la \vec{s_n}|,
\label{S_Pauli}
\end{eqnarray}
where $\hat{X}$ stands for the Pauli $\sigma_x$ operator. This structure of $\hat{S}$ reflects the property of the hypercube that moving along an edge from $\vec{v}$ corresponds to flipping one bit of $\vec{v}$. This structure is also useful in determining its group of symmetries as we shall see below.

An {\it automorphism} of a graph is a permutation of its vertices such that it leaves the graph unchanged. The set of all such permutations is the {\it automorphism group} of the graph. When the edge labels or colors in the graph are important, as in the case of a discrete quantum walk, we restrict ourselves to those automorphisms which preserve the edge labels. In other words, an edge connecting two vertices has the same label before and after the permutation. Such automorphisms are called {\it direction-preserving}. In general, we could consider automorphisms where we permute the direction labels along with the vertices to obtain the same graph with the same coloring. This would form a larger group $G$ of which the direction-preserving automorphisms are a subgroup $H$.

Since the vertex Hilbert space $\mathcal{H}^v$ has its basis elements in one-to-one correspondence with the vertices of the graph, and the coin Hilbert space has a basis in correspondence with the direction labels, the automorphisms (which are just permutations of vertices and directions) are permutation matrices. In fact, these are all the permutation matrices on $\mathcal{H}^v\otimes\mathcal{H}^c$ that leave $\hat{S}$ unchanged, i.e., $\{$all $\hat{P}\, |\, \hat{P}\hat{S}\hat{P}=\hat{S}$, where $\hat{P}$ is a permutation matrix$\}$. In this representation, any direction-preserving automorphism has the structure $\hat{P}_v\otimes \hat{I}_c$, where $\hat{P}_v$ acts solely on $\mathcal{H}^v$ and $\hat{I}_c$ on $\mathcal{H}^c$. Such automorphisms become important if we wish to consider the symmetries of $\hat{U} \equiv  \hat{S}(\hat{I}\otimes\hat{C})$. Clearly, any automorphism of this type is a symmetry of $\hat{U}$, since
\begin{equation}
(\hat{P}_v\otimes \hat{I}_c) \left[ \hat{S}(\hat{I}\otimes\hat{C}) \right] (\hat{P}_v\otimes \hat{I}_c) = \left[ (\hat{P}_v\otimes \hat{I}_c)\hat{S}(\hat{P}_v\otimes \hat{I}_c) \right] (\hat{I}\otimes\hat{C}) = \hat{S}(\hat{I}\otimes\hat{C}) .
\end{equation}
Elements of $G$ in general do not act trivially on the coin space. Because of this, they need not be symmetries of $\hat{U}$ unless the coin flip operator $\hat{C}$ respects these symmetries. 

To illustrate all this, consider the example of a hypercube in 2 dimensions (i.e., a square). The vertex labels are $\{(00),(01),(10),(11)\}$ (which also form a basis for $\mathcal{H}^v$); the edges connecting $(00)$ to $(01)$ and $(10)$ to $(11)$ are both labeled $1$, and the edges connecting $(00)$ to $(10)$ and $(01)$ to $(11)$ are both labeled $2$. Thus, the transformation $(00)\leftrightarrow (01)$ and $(10) \leftrightarrow (11)$, or the transformation $(00)\leftrightarrow (10)$ and $(01) \leftrightarrow (11)$, or both together, are automorphisms of this graph which need no permutation of the directions. Together with the identity automorphism (which permutes nothing), these permutations form the direction-preserving subgroup $H$. In a matrix representation on the Hilbert space $\mathcal{H}^v\otimes\mathcal{H}^c$, they are,
\begin{equation}\nonumber
\begin{pmatrix}
0 & 1 & 0 & 0\\
1 & 0 & 0 & 0\\
0 & 0 & 0 & 1\\
0 & 0 & 1 & 0
\end{pmatrix}\otimes \hat{I}_c,
\begin{pmatrix}
0 & 0 & 1 & 0\\
0 & 0 & 0 & 1\\
1 & 0 & 0 & 0\\
0 & 1 & 0 & 0
\end{pmatrix}\otimes \hat{I}_c,
\begin{pmatrix}
0 & 0 & 0 & 1\\
0 & 0 & 1 & 0\\
0 & 1 & 0 & 0\\
1 & 0 & 0 & 0
\end{pmatrix}\otimes \hat{I}_c,
\begin{pmatrix}
1 & 0 & 0 & 0\\
0 & 1 & 0 & 0\\
0 & 0 & 1 & 0\\
0 & 0 & 0 & 1
\end{pmatrix}\otimes \hat{I}_c
\end{equation}
where $\hat{I}_c$ is the $2\times 2$ identity matrix acting on the coin space. These permutations can be easily seen to be $H=\{\hat{I}\otimes\hat{X}\otimes\hat{I}, \hat{X}\otimes\hat{I}\otimes\hat{I}, \hat{X}\otimes\hat{X}\otimes\hat{I}, \hat{I}\otimes\hat{I}\otimes\hat{I} \}$.  Just as in the representation of $\hat{S}$ matrix in terms of the Pauli $\hat{X}$ operators given by Eq.~(\ref{S_Pauli}), this group denotes a bit flip in the first, second or both bits of each vertex, together with the identity, which gives no flip.  (See Fig.~2.)

\begin{figure}[t]
\begin{center}
\includegraphics[scale=0.8]{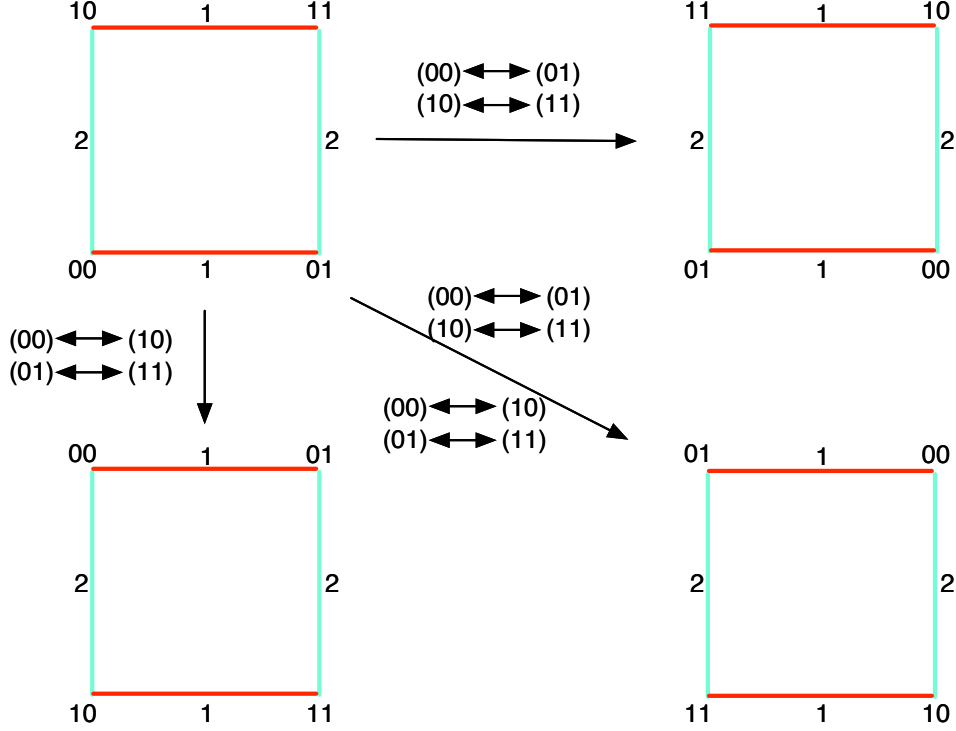}
\end{center}
\caption{(Color online)The direction-preserving automorphism group of the n=2 hypercube.} \label{fig2}
\end{figure}

\begin{figure}[t]
\begin{center}
\includegraphics[scale=0.8]{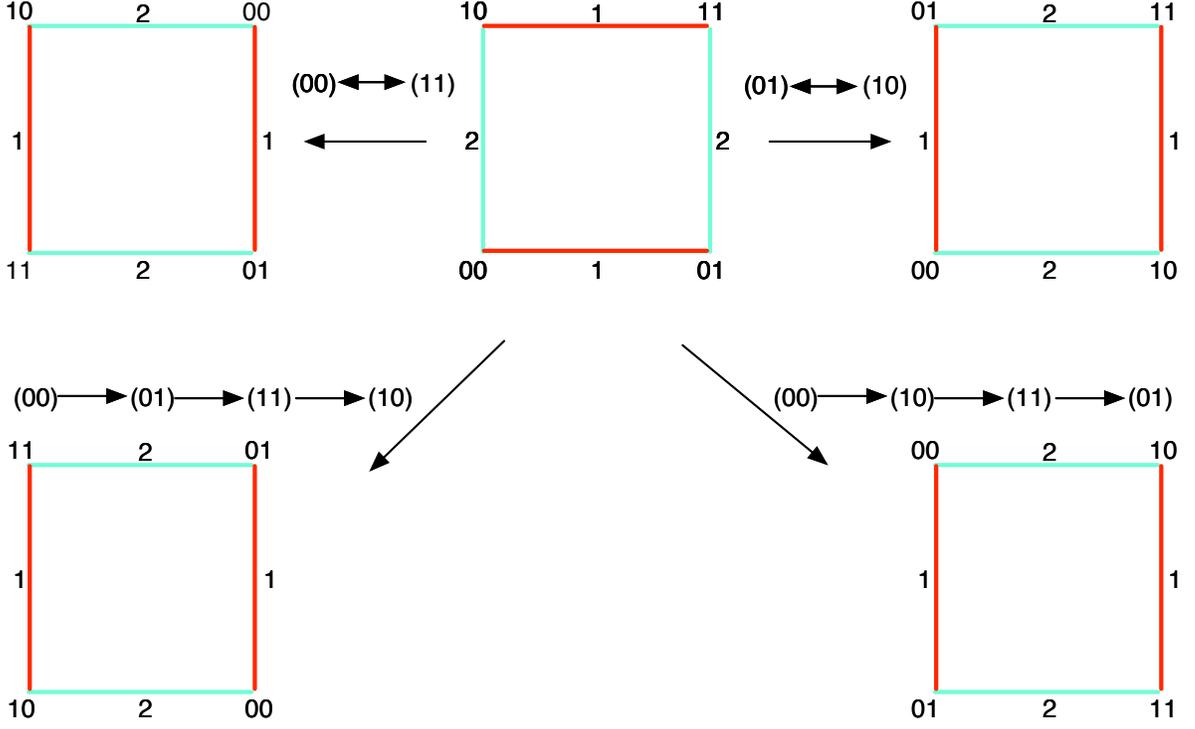}
\end{center}
\caption{(Color online) Automorphisms which interchange directions for the n=2 hypercube.} \label{fig3}
\end{figure}

The permutation $(10)\leftrightarrow (01)$, reflecting along the diagonal while keeping $(00)$ and $(11)$ fixed, will be an automorphism only if we interchange the directions $1\leftrightarrow 2$. Similarly, the permutations $(00)\leftrightarrow (11)$, $(00)\rightarrow (01)\rightarrow (11)\rightarrow (10)$ and $(00)\rightarrow (10)\rightarrow (11)\rightarrow (01)$ are automorphisms when we interchange the two directions. If we view these permutations along with those obtained above, we obtain a new group $G$ for which $H$ is a subgroup. In a matrix representation, the new automorphisms are,
\begin{equation}\nonumber
\begin{pmatrix}
0 & 1 & 0 & 0\\
0 & 0 & 1 & 0\\
0 & 0 & 0 & 1\\
1 & 0 & 0 & 0
\end{pmatrix}\otimes \hat{X}_c,
\begin{pmatrix}
1 & 0 & 0 & 0\\
0 & 0 & 0 & 1\\
0 & 0 & 1 & 0\\
0 & 1 & 0 & 0
\end{pmatrix}\otimes \hat{X}_c,
\begin{pmatrix}
0 & 0 & 1 & 0\\
0 & 1 & 0 & 0\\
1 & 0 & 0 & 0\\
0 & 0 & 0 & 1
\end{pmatrix}\otimes \hat{X}_c,
\begin{pmatrix}
0 & 0 & 0 & 1\\
1 & 0 & 0 & 0\\
0 & 1 & 0 & 0\\
0 & 0 & 1 & 0
\end{pmatrix}\otimes \hat{X}_c,
\end{equation}
where $\hat{X}_c$ acts on the coin space and corresponds to an interchange of the two directions.  (See Fig.~3.)  These four elements of $G$ need not be symmetries of $\hat{U}$, since the coin need not be symmetric under conjugation with $\hat{X}_c$. However, for the hypercube, if we use the Grover diffusion matrix as the coin, then automorphism group $G$ is indeed its group of symmetries, since the Grover coin is symmetric under any permutation of its basis elements. The symmetry group of the evolution operator would be $H$ if the DFT coin is used, since the DFT does not have permutation symmetry. It is important to note that the symmetry group defined above is not the only thing that influences the degeneracy of the evolution operator. Degeneracy of the coin flip operator $\hat{C}$ can also induce degeneracy in the evolution operator, and the coin may be degenerate even if it does not have permutation symmetry (like the DFT coin). We discuss this in more detail in Sec. IV B after describing the relationship between symmetry and degeneracy.

It can be shown that the direction-preserving automorphism group $H$ for any Cayley graph is isomorphic to the group on which the graph is defined. This is because any direction-preserving automorphism of a Cayley graph is a left translation by a  group element, and conversely all left translations are direction-preserving automorphisms. The first part of the statement is easy to see. Consider any left translation $L_a:G\rightarrow G$ which has the action $L_a(g)=a g$, for all $g\in G$. Now, given vertices $g$ and $h$ in $G$, they are connected by an edge from $g$ to $h$ if $g^{-1}h=s$, where $s\in S$. Clearly, after the transformation we still have $(a g)^{-1}(ah)=g^{-1}h=s$ and hence this automorphism preserves the direction labels. Since the group elements are basis states of the vertex Hilbert space, a left translation by a group element corresponds to a permutation matrix on this Hilbert space. The fact that every direction-preserving automorphism of a Cayley graph is a left translation by a group element becomes important in the discussion of regular and irreducible representations in Sec. IV.

Before we conclude this section, let us explicitly construct the representation of the automorphism group for the hypercube, which has $H\cong\mathcal{Z}_2^n$ and $G\cong H\cdot S_n$. In terms of the Pauli operators the representation of $H$ is $\{\hat{I}\hat{I}\hat{I}\cdots \hat{I}\otimes\hat{I}_c, \hat{X}\hat{I}\hat{I}\cdots \hat{I}\otimes\hat{I}_c, \hat{X}\hat{X}\hat{I}\cdots \hat{I}\otimes\hat{I}_c,\dots,\hat{X}\hat{X}\hat{X}\cdots \hat{X}\otimes\hat{I}_c\}$, where the tensor product symbol has been dropped in the vertex space, and $\hat{I}_c$ is the identity operator in the coin space. In fact, the representation of $H$ for any Cayley graph will be of the form $\hat{P}\otimes \hat{I}$, where $\hat{P}$ is a permutation matrix on the vertex space and $\hat{I}$ is the identity on the coin space. The group $G$ will become $H\cdot S_n=\{h \cdot \pi | h\in H, \pi\in S_n\}$, where $S_n$ is the permutation group on $n$ elements which is assumed to act on the direction labels.

\section{Hitting times}

\subsection{Definition for quantum walks}

The hitting time $\tau_h$ of a classical random walk is defined as the average time for the walk to hit a designated `final' vertex $v_f$ given that the walk began with some initial distribution $p_i$:
\begin{equation}\label{ht1}
\tau_h = \sum_{t=0}^\infty t p(t),
\end{equation}
where $p(t)$ is the probability of being in the final vertex for the first time at time step $t$. In order to carry this notion of hitting time over to the quantum case, we need to make the meaning of $p(t)$ more precise. In particular, we need to define clearly what ``for the first time'' means for a quantum walk.  As described in \cite{KroviBrun}, we do this by performing a measurement of the particle at every step of the walk to see if the particle has reached the final vertex or not. The measurement $M$ which is used has projectors $\hat{P}_f$ and $\hat{Q}_f=\hat{I}-\hat{P}_f$ representing the particle being found or not found at the final vertex, respectively.  The projector is defined $\hat{P}_f=|x_f\ra\la x_f|\otimes\hat{I}_c$, where $|x_f\ra$ is the final vertex state and $\hat{I}_c$ is the identity operator on the coin space. Using this definition, each step of the measured walk consists of an application of the unitary evolution operator $\hat{U}$ followed by the measurement $M$.

Now we can use the same expression (\ref{ht1}) for the hitting time, where the probability $p(t)$ becomes
\begin{equation}\label{prob.eqn}
p(t)=\tr\{\hat{P}_f\hat{U}[\hat{Q}_f\hat{U}]^{t-1}
\rho_0[\hat{U^{\dag}}\hat{Q}_f]^{t-1}\hat{U^{\dag}}\hat{P}_f\} .
\end{equation}
To explicitly sum the series (\ref{ht1}) using the expression for $p(t)$ in Eq. (\ref{prob.eqn}), we rewrite the expression in terms of {\it superoperators} (linear transformations on operators) $\mathcal{N}$ and $\mathcal{Y}$, defined by
\begin{eqnarray}\label{superops}
\mathcal{N}\rho = \hat{Q}_f\hat{U}\rho\hat{U^{\dag}}\hat{Q}_f \nonumber\\
\mathcal{Y}\rho=\hat{P}_f\hat{U}\rho\hat{U^{\dag}}\hat{P}_f.
\end{eqnarray}
In terms of $\mathcal{N}$ and $\mathcal{Y}$,
$p(t)=\tr\{\mathcal{Y}\mathcal{N}^{t-1}\rho_0\}$.
We introduce a new superoperator $\mathcal{O}(l)$ which depends on a real parameter $l$:
\begin{equation}
\mathcal{O}(l) = l\sum_{t=1}^\infty (l\mathcal{N})^{t-1} ,
\label{superop_sum}
\end{equation}
which is a function of a parameter $l$.  The hitting time now becomes
\begin{equation}
\tau_h = \frac{d}{dl} \tr\{ \mathcal{Y} \mathcal{O}(l) \rho_0 \} \biggr|_{l=1}.
\label{derivative_form}
\end{equation}

If the superoperator $\mathcal{I}-l\mathcal{N}$ is invertible, then we can replace the sum (\ref{superop_sum}) with the closed form
\begin{equation}
\mathcal{O}(l) = l(\mathcal{I}-l\mathcal{N})^{-1}.
\end{equation}
(The case when $\mathcal{I}-l\mathcal{N}$ is not invertible is discussed in detail in the next section.)   The derivative in (\ref{derivative_form}) is
\begin{equation}
\frac{d\mathcal{O}}{dt}(1)
= (\mathcal{I}-\mathcal{N})^{-1}+\mathcal{N}(\mathcal{I}-\mathcal{N})^{-2}
= (\mathcal{I}-\mathcal{N})^{-2}.
\end{equation}
This gives us the following expression for the hitting time:
\begin{equation}
\tau_h = \tr\{\mathcal{Y}(\mathcal{I}-\mathcal{N})^{-2}\rho_0\}.
\label{closed_form_tau}
\end{equation}


To evaluate (\ref{closed_form_tau}), we write these superoperators as matrices using Roth's lemma \cite{roth}. As shown in \cite{KroviBrun}, we can then {\it vectorize} the density operators and operators on states, and write the action of superoperators as simple matrix multiplication.  Any matrix can be vectorized by turning its rows into columns and
stacking them up one by one, so that a $D\times D$ matrix becomes
a column vector of size $D^2$.  For example:
\begin{equation}
\begin{pmatrix}
a_{11} & a_{12} & a_{13} \\
a_{21} & a_{22} & a_{23} \\
a_{31} & a_{32} & a_{33}
\end{pmatrix}
\rightarrow
\begin{pmatrix}
a_{11} \\
a_{12} \\
a_{13} \\
a_{21} \\
a_{22} \\
a_{23} \\
a_{31} \\
a_{32} \\
a_{33}
\end{pmatrix} . \nonumber
\end{equation}
Consequently the superoperators become matrices
of size $D^2\times D^2$. This method of vectorization takes operators on one Hilbert space $\mathcal{H}$ to vectors in another Hilbert space $\mathcal{H}'=\mathcal{H}\otimes\mathcal{H}^\ast$ and so superoperators in $\mathcal{H}$ are operators in $\mathcal{H}'$. Note that a basis $\{|u_{i j}\ra\}$ for $\mathcal{H}'$ can be obtained from a basis $\{|v_i\ra\}$ for $\mathcal{H}$ by defining
\begin{equation}
|u_{i j}\ra=|v_i\ra\otimes |v_j\ra^\ast .
\end{equation}
For our superoperators $\mathcal{N}$ and $\mathcal{Y}$ we then get
\begin{eqnarray}
(\mathcal{N}\rho)^v &=& \left[ (\hat{Q}_f\hat{U})\otimes(\hat{Q}_f\hat{U})^\ast \right] \rho^v ,
\nonumber \\
(\mathcal{Y}\rho)^v &=& \left[ (\hat{P}_f\hat{U})\otimes(\hat{P}_f\hat{U})^\ast \right] \rho^v .
\end{eqnarray}
Let $\mathbf{N}=(\hat{Q}_f\hat{U})\otimes(\hat{Q}_f\hat{U})^\ast$ and $\mathbf{Y}=(\hat{P}_f\hat{U})\otimes(\hat{P}_f\hat{U})^\ast$. The hitting time becomes
\begin{equation}\label{hittime}
\tau_h = I^v \cdot \left( \mathbf{Y}(\mathbf{I}-\mathbf{N})^{-2}\rho^v \right) .
\end{equation}
Using this vectorization transformation, we treat the superoperators as operators on a larger Hilbert space and thus can find their inverses. But, the expression in Eq. (\ref{hittime}) is not always well defined, as we shall see below, because the matrix $\mathbf{I}-\mathbf{N}$ may not be invertible. We will show that when it is not invertible, it means that the hitting time becomes infinite for some initial states, and vice versa. This property of quantum walks having infinite hitting times does not have a classical analogue.
 
\subsection{Infinite hitting times}

We will now show that it is possible for hitting times to be {\it infinite}, and derive a sufficient condition for the unitary evolution operator to allow infinite hitting times. We begin by forming the projector $\hat{P}$ onto the subspace spanned by all eigenstates of $\hat{U}$ which have no overlap with the final vertex.  This projector is orthogonal to the projector onto the final vertex, $\hat{P}\hat{P}_f = \hat{P}_f\hat{P} = 0$, and commutes with $\hat{U}$, $[\hat{U},\hat{P}]=0$.  We assume (for the moment) that this projector is nonzero; later, we will find a sufficient condition for this to be true, and exhibit quantum walks which satisfy this condition.  We can write any initial state as a superposition of a state in the subspace projected onto by $\hat{P}$ and a state orthogonal to it, giving the decomposition
\begin{equation}
|\Psi\ra=\hat{P} |\Psi\ra + (\hat{I}-\hat{P}) |\Psi\ra .
\end{equation}
It is easy to see that if $|\Psi\ra$ lies entirely inside $\hat{P}$, i.e., $\hat{P}|\Psi\ra=|\Psi\ra$, then under the evolution the subsequent states will never have any component in the final state and the probability defined in eq. (\ref{prob.eqn}) will be zero. Indeed, since $[\hat{P},\hat{U}]=0$ and $[\hat{P},\hat{Q}_f]=0$,
\begin{eqnarray}
p(t) &=& \tr\{\hat{P}_f\hat{U}[\hat{Q}_f\hat{U}]^{t-1}
\rho_0[\hat{U^{\dag}}\hat{Q}_f]^{t-1}\hat{U^{\dag}}\hat{P}_f\}\\
&=& \tr\{\hat{P}_f\hat{U}[\hat{Q}_f\hat{U}]^{t-1}
\hat{P}\rho_0\hat{P}[\hat{U^{\dag}}\hat{Q}_f]^{t-1}\hat{U^{\dag}}\hat{P}_f\}\\
&=& \tr\{\hat{P}_f\hat{P}\hat{U}[\hat{Q}_f\hat{U}]^{t-1}
\rho_0[\hat{U^{\dag}}\hat{Q}_f]^{t-1}\hat{U^{\dag}}\hat{P}^{\dag}\hat{P}_f\}\\
&=& 0 ,
\end{eqnarray}
where $\rho_0=|\Psi\ra\la\Psi|$. Therefore, the hitting time for this initial state is infinite. More generally, if $|\Psi\ra$ has nonzero overlap with $\hat{P}$, $\hat{P}|\Psi\ra \ne 0$, then that component of $|\Psi\ra$ can never reach the final vertex. The probability of ever hitting the final vertex if one starts with this initial state is
\begin{equation}
p=|\la\Psi|(\hat{I}-\hat{P})|\Psi\ra|^2 < 1,
\end{equation}
and the hitting time is again infinite.

To construct this projector, we look at the spectral decomposition of $\hat{U}$. If $\hat{U}$ has at least one sufficiently degenerate eigenspace, then we can construct a subspace of this eigenspace which has a zero overlap with the final state. For instance, consider one such degenerate eigenspace which has a degeneracy of $k$. Since the vector space at the final vertex is $d$ dimensional (i.e., it has $d$ coin degrees of freedom), we would be solving the following $d\times k$ system of homogeneous equations:
\begin{equation}
a_1\begin{pmatrix}
v_{N-d+1}^1\\
\vdots\\
v_N^1\end{pmatrix}+
a_2\begin{pmatrix}
v_{N-d+1}^2\\
\vdots\\
v_N^2\end{pmatrix}+
\cdots+
a_k\begin{pmatrix}
v_{N-d+1}^k\\
\vdots\\
v_N^k\end{pmatrix}=0 .
\end{equation}
Here, we use a labeling where the final vertex in some coin state occupies the last $d$ entries of the eigenvectors. The subscript refers to the component of the eigenvector, and the superscript distinguishes the eigenvectors in the degenerate eigenspace. This system is under-determined if $k>d$, and it will always have a nontrivial solution---in fact, it will have a space of solutions of dimension $k-d$. Therefore, it is sufficient that there exist at least one eigenspace of $\hat{U}$ with dimension greater than the dimension of the coin, in order to have a nonzero projector $\hat{P}$.  If there is more than one degenerate eigenvalue with multiplicity greater than $d$, the subspace projected onto by $\hat{P}$ will include {\it all} the eigenvectors of $\hat{U}$ which have no overlap with the final vertex.

The condition derived above is closely related to the question of invertibility of $\mathbf{I}-\mathbf{N}$ in Eq.~(\ref{hittime}) of the previous section. Here we show that $\mathbf{I}-\mathbf{N}$ is not invertible if and only if the projector $\hat{P}$ is nonzero. Furthermore, in the case when $\mathbf{I}-\mathbf{N}$ is not invertible, the hitting time for a state or density matrix whose support has no overlap with $\hat{P}$ is calculated by replacing the inverse of $\mathbf{I}-\mathbf{N}$ with its pseudo-inverse in Eq.~(\ref{hittime}).

Assume that the projector $\hat{P}$ is nonzero. Then there is at least one eigenvector $|v\ra$ of $\hat{U}$ such that $\hat{P}_f |v\ra=0$. Therefore,
\begin{equation}
(\hat{Q}_f \hat{U}\otimes \hat{Q}_f^\ast \hat{U}^\ast) (|v\ra\otimes |v\ra^\ast) = (\hat{Q}_f \otimes \hat{Q}_f^\ast)  (|v\ra\otimes |v\ra^\ast) = |v\ra\otimes |v\ra^\ast ,
\end{equation}
since $\hat{U} |v\ra=\exp(i\theta) |v\ra$ and $\hat{U}^\ast |v\ra^\ast = \exp(-i\theta) |v\ra^\ast$, and $\hat{P}_f |v\ra=0$ implies that $\hat{Q}_f |v\ra = |v\ra$. Therefore, $(\mathbf{I}-\mathbf{N})|v\ra\otimes |v\ra^\ast=0$, $\mathbf{I}-\mathbf{N}$ has a nonzero nullspace, and hence is not invertible.  This proves the ``if'' direction.

To prove the ``only if'' direction, assume that $\mathbf{I}-\mathbf{N}$ is not invertible. This implies that there exists a normalized vector $|u\ra \in {\cal H}\otimes{\cal H}^\ast$ such that
\begin{equation}
(\hat{I}-\hat{Q}_f \hat{U}\otimes \hat{Q}_f^\ast \hat{U}^\ast) |u\ra = 0 \Longrightarrow
\hat{Q}_f \hat{U}\otimes \hat{Q}_f^\ast \hat{U}^\ast |u\ra=|u\ra .
\label{N_nullspace}
\end{equation}
The vector $|u\ra$ is an eigenvector of $\hat{Q}_f \hat{U}\otimes \hat{Q}_f^\ast \hat{U}^\ast$ with eigenvalue 1.  Since $\hat{Q}_f\otimes\hat{Q}_f^\ast$ is a projector, the vector $|u\ra$ must therefore lie in the eigenspace of eigenvalue $1$ of both $\hat{Q}_f\otimes\hat{Q}_f^\ast$ and $\hat{U}\otimes\hat{U}^\ast$.  This can only be true if $|u\ra$ is of the form:
\begin{equation}
|u\ra=\sum_{i,j}a_{i j}|v_i\ra\otimes |v_j\ra^\ast ,
\end{equation}
where the $\{|v_i\ra\}$ are eigenvectors of $\hat{U}$, and $a_{ij}$ is only nonzero if $|v_i\ra$ and $|v_j\ra$ lie in the same eigenspace of $\hat{U}$ and $\hat{P}_f|v_i\ra = \hat{P}_f|v_j\ra = 0$. Note that the vector $|u\ra$ need not correspond to a physical state in the Hilbert space of the walk. But the existence of such a $|u\ra$ means that the projector $\hat{P}$ is nonzero, since there must exist at least one $|v_i\ra$ which has a zero overlap with $\hat{P}_f$.  This proves the ``only if'' direction.

We have shown that if $\mathbf{I}-\mathbf{N}$ is not invertible then the projector $\hat{P}$ is nonzero and vice-versa. Now we will see that if a density operator $\rho$ is orthogonal to $\hat{P}$, $\rho\hat{P} = \hat{P}\rho = 0$, its corresponding vector $\rho^v$ lies outside the null space of $\mathbf{I}-\mathbf{N}$. If $\rho$ is orthogonal to $\hat{P}$, then when written in the eigenbasis $\{|v_i\ra\}$ of $\hat{U}$, the diagonal components $\la v_i|\rho|v_k\ra$ are nonzero if and only if $\hat{P}_f|v_i\ra\neq 0$. This implies that for the corresponding vectorized quantity $\rho^v$, we have 
\begin{equation}
I^v.(\hat{P}_f\otimes\hat{P}_f^\ast)\rho^v\neq 0 . 
\end{equation}
For any $\rho^v$,
\begin{equation}
I^v.(\hat{P}_f\otimes\hat{P}_f^\ast) (\hat{Q}_f\hat{U}\otimes\hat{Q}_f^\ast\hat{U}^\ast) \rho^v = 0 .
\end{equation}
Therefore, $\mathbf{N}\rho^v \ne \rho^v$ i.e., $\rho^v$ does not lie in the null space of $\mathbf{I}-\mathbf{N}$. Moreover, if $\rho$ is orthogonal to $\hat{P}$, then so are $\mathcal{N}\rho$ and $\mathcal{Y}\mathcal{N}^t\rho$. This is easy to see, since from Eq.~(\ref{superops}) we get
\begin{equation}
\hat{P}\mathcal{N}\rho = \hat{P}(\hat{Q}_f\hat{U}\rho\hat{U}^{\dag}\hat{Q}_f) = \hat{Q}_f\hat{U}\hat{P}\rho\hat{U}^{\dag}\hat{Q}_f = 0 ,
\end{equation}
since $\hat{P}\hat{Q}_f=\hat{P}$ and $[\hat{U},\hat{P}]=0$. Similarly, we obtain $\hat{P}\mathcal{Y}\rho = 0$, since
\begin{equation}
\hat{P}\mathcal{Y}\rho=\hat{P}(\hat{P}_f\hat{U}\rho\hat{U}^{\dag}\hat{P}_f)=0 .
\end{equation}
Therefore, if $\rho$ is orthogonal to $\hat{P}$, then all the terms of the type $\mathcal{Y}\mathcal{N}^{t-1}\rho$ for all $t$ and hence all the terms inside the trace of Eq. (\ref{derivative_form}) are orthogonal to $\hat{P}$, which means that the vectorized versions of all terms inside the trace in Eq.~(\ref{derivative_form}) lie outside the null space of $\mathbf{I}-\mathbf{N}$. Thus, for states that do not overlap with $\hat{P}$, the hitting time is finite as would be expected, and is given by the same formula Eq.~(\ref{hittime}) with the inverse replaced by a pseudo-inverse. 

We focus on the discrete time quantum walk for the remainder of this section, considering the walk on the hypercube in particular. It was observed in numerical simulations \cite{KroviBrun} that for a walk on the hypercube with the DFT coin, an initial state given by
\begin{equation}\label{init.state}
|\Psi\ra=|00\cdots 0\ra\otimes \frac{1}{\surd{d}}\sum_i|i\ra
\end{equation}
has an infinite hitting time. The phenomenon of infinite hitting times is not restricted to the walk with the DFT coin, however.  Numerical simulations, followed by analytical calculations, have shown that it also occurs with the Grover coin, but for different initial states. In fact, it turns out that for the Grover coin, the symmetric initial state (\ref{init.state}) is the {\it only} initial state localized at the vertex $|00\cdots0\ra$ that has a {\it finite} hitting time. Any other superposition of the coin states for that vertex will give an {\it infinite} hitting time, because all such states have a nonzero overlap with $\hat{P}$.

Given any vertex $v$ on a graph, it is natural to ask if there exists any superposition of its coin states which overlaps with $\hat{P}$, and for which coin state the overlap is maximum and for which it is minimum (or zero). We can write the projector $\hat{P}$ in the form
\begin{equation}
\hat{P}=\sum_{i,j,k,l} A_{ijkl}|x_i\ra\la x_j|\otimes |k\ra\la l|,
\end{equation}
where $\{|x_i\ra\}$ are the vertices and $\{|k\ra\}$ are the directions. Suppose the initial state is
\begin{equation}
|\Psi\ra=|v\ra\otimes\sum_i\alpha_i|i\ra \equiv |v\ra \otimes |\alpha\ra.
\end{equation}
Its overlap with the projector $\hat{P}$ is given by
\begin{equation}
\la\Psi|\hat{P}|\Psi\ra=\sum_{k,l} A_{vvkl} \alpha_k^*\alpha_l .
\end{equation}
To find the superposition of coin states such that the overall initial state has the least (or greatest) overlap with $\hat{P}$, define the matrix
\begin{equation}
\hat{C}_v = \tr_{\rm vertices}\left\{\hat{P}(|v\ra\la v|\otimes\hat{I}_{\rm coin})\right\} ,\ \ (\hat{C}_v)_{kl} = A_{vvkl} .
\end{equation}
The overlap of the initial state with $\hat{P}$ can be written in terms of this matrix as
\begin{equation}
\la\Psi|\hat{P}|\Psi\ra=\la\alpha|\hat{C}_v|\alpha\ra .
\end{equation}
The matrix $\hat{C}_v$ is Hermitian and positive, and hence has a spectral decomposition into a complete orthonormal basis of eigenstates with non-negative eigenvalues.  Assuming that $\{\lambda_i,|e_i\ra\}$ is the spectral decomposition of $\hat{C}_v$, we can rewrite the overlap as
\begin{equation}
\la\Psi|\hat{P}|\Psi\ra=\sum_i \lambda_i |\la\alpha|e_i\ra|^2 .
\end{equation}
From the above expression, we see that the overlap is maximum (or minimum) if $|\alpha\ra$ is in the direction of the eigenvector with the largest (or smallest) eigenvalue, and zero if $|\alpha\ra$ is an eigenvector with zero eigenvalue. Therefore, if $\hat{C}_v$ does not have a zero eigenvalue (i.e., is positive definite), then for that vertex every superposition of coin states will overlap with $\hat{P}$.  In other words, the hitting time will be infinity if one starts at that vertex no matter what coin state one chooses. Numerical calculations for the Grover and the DFT coins on the hypercube show that for the vertex $|00\cdots 0\ra$, an equal superposition of coin states is the only superposition that has a zero overlap with $\hat{P}$ for the Grover coin, and no superposition of coin states has a zero overlap for the DFT coin. Moreover, $\tr\{\hat{P}\}$ for the Grover coin on the hypercube for $n=4$ is $32$, which is fully half the dimension of the total space (dim=$2^4\cdot4=64$). These examples suggest that infinite hitting times may be a generic phenomenon on graphs with symmetry.

\section{Symmetries and degeneracy}

\subsection{Representations of finite groups}

We saw in the previous section that degeneracy of the evolution operator leads to infinite hitting times. One of the main sources of degeneracy in quantum mechanics is {\it symmetry}. Since the $\hat{S}$ matrix which makes up part of the evolution operator encodes the connections of the graph, it is natural to expect that symmetries of the graph will produce symmetries of the evolution operator.  In this section, we analyze this idea.

Is it sufficient to consider only the symmetries of the graph?  Apart from the symmetries induced from the graph, the evolution operator may have additional symmetries of its own which lead to additional degeneracy. But such symmetries are difficult to analyze in generality, and depend on the details of how one defines the quantum walk. First, there is a choice between the discrete and the continuous walk. Second, for the discrete walk, if the structure $\hat{U}=\hat{S}(\hat{I}\otimes \hat{C})$ that was described Sec.~II is used, there is still the freedom to use any unitary matrix as the coin.  In order to generalize our discussion of symmetries to any walk, we restrict attention to the symmetries induced by the graph alone. This naturally leads us to the question:  ``Are there graphs with sufficient symmetry such that, for any walk that is defined on the graph, the resultant evolution operator will have enough degeneracy to give rise to infinite hitting times?''  It turns out that such graphs do exist, and we give an example of such a class of graphs.  We will comment briefly on the effect of additional symmetries of the coin for the hypercube later in the paper.

First, let us briefly review linear representations of finite groups, and describe how the symmetry group of $\hat{U}$ affects its degeneracy by determining the dimensions of the irreducible representations of the group.  (For further details, see for example \cite{Sternberg}.)  A linear representation of a finite group $G$ on a finite-dimensional vector space $V$ is a map $\sigma :G \rightarrow GL(V)$ , such that $\sigma(g h)=\sigma(g)\sigma(h)$.  $GL(V)$ is the space of invertible linear maps of $V$ onto itself.  If $V$ is $d$-dimensional, and we choose a basis of $d$ vectors in $V$, then $\sigma (g)$ for $g\in G$ becomes a $d\times d$ invertible matrix. The trace of this matrix is called the {\it character} of the representation. Therefore, characters are maps $\chi: G\rightarrow \mathbb{C}$ with $\chi(g)=\tr(\sigma(g))$.  Two representations $\sigma_1$ and $\sigma_2$ of the group $G$ on vector spaces $V_1$ and $V_2$, respectively, are considered equivalent if there exists an invertible linear map $\tau :V_1\rightarrow V_2$ such that $\tau\circ\sigma_1(g)=\sigma_2(g)\circ\tau$, for all $g\in G$. The characters of equivalent representations are equal, a fact which follows from the cyclic property of the trace operator.

Assume that the vector space $V$ has an inner product defined on it. Since the group $G$ is assumed finite, it can be shown that any representation $\sigma$ is equivalent to a {\it unitary} representation---there exists a basis for $V$ in which $\sigma(g)$ is a unitary matrix for all $g\in G$ (see \cite{Serre}). A vector space $W$ is said to be {\it invariant} or {\it stable} under the action of $G$ if $x\in W\Rightarrow \sigma(g)x\in W$ for all $g\in G$. If the vector space $V$ has a subspace $W$ which is invariant under the action of $G$, then it can be shown that its orthogonal complement $W^\perp$ is also invariant under $G$, and so $V$ can be decomposed
\begin{equation}
V=W\oplus W^\perp .
\end{equation}
This means that the representation $\sigma$ as a matrix on $V$ can be written in a block diagonal form consisting of two blocks as
\begin{equation}
\sigma=\begin{pmatrix}
\sigma |_W & 0\\
0 & \sigma |_{W^\perp}
\end{pmatrix} ,
\end{equation}
where $\sigma |_W$ and $\sigma |_{W^\perp}$ are the restrictions of $\sigma$ to the subspaces $W$ and $W^\perp$. 

The linear map $\sigma: G\rightarrow GL(V)$ is called an {\it irreducible} representation (irrep), if it is a representation and no non-trivial subspace of $V$ is stable under the action of $G$.  Equivalently, it is an irreducible representation if it is not a direct sum of two representations. {\it Any} representation on $V$ can, by an appropriate choice of basis, be written as a representation in block diagonal form, where each block corresponds to an irrep.  So $V$ can be decomposed in the following way:
\begin{equation}\label{V_decomp}
V=W_1\oplus W_2\oplus\cdots\oplus W_k ,
\end{equation}
where each of the $W_i$ is stable under the action of $\sigma(G)$.  This decomposition is unique, up to reordering of the spaces and an overall equivalence transformation.  By an abuse of notation, we will use the same labels ($V$ or $W_i$) to refer both to the vector space and to the group representation on that space. Each component $W_i$ of the decomposition in (\ref{V_decomp}) is isomorphic to an irrep, and the number of such $W_i$ isomorphic to a given irrep does not depend on the details of the decomposition.

If we define an inner product for characters,
\begin{equation}
(\phi|\chi)=(1/|G|)\sum_{g\in G} \phi(g)\chi(g)^\ast ,
\end{equation}
it can be shown that the number of times an irrep with character $\chi_i$ occurs in a representation with a character $\chi$ is given by $(\chi|\chi_i)$ (\cite{Sternberg}).

We now define a {\it regular} representation of a group $G$. Suppose $|G|=n$, and let $V$ be an $n$-dimensional vector space. Define a basis for V $\{e_t\}$ which is labeled by the group elements of $t\in G$. A regular representation of $G$ is a map $\sigma :G\rightarrow V$ such that $\sigma(g) e_t = e_{gt}$. It can easily be seen that the action of $G$ on the basis vectors is a left translation, and in matrix form the representations of the group elements will be permutation matrices. An important property of the regular representation is that its decomposition into irreps contains {\it all} the irreps of the group in it, and each irrep has a multiplicity equal to its dimension (\cite{Sternberg}). Therefore, we can write
\begin{equation}
V=n_1W_1\oplus n_2W_2\oplus\cdots\oplus n_kW_k ,
\end{equation}
where we use the notation $nW$ to mean $n$ copies of the space $W$, $nW = W\oplus W\oplus\cdots W$.  The $\{W_i\}$ are all the inequivalent irreps of the group, and $n_i$ is the dimension of $W_i$.

It will be useful to note that all Abelian groups have one-dimensional irreps. The converse of this statement is also true:  all groups which have only one-dimensional irreps are Abelian. Finally, suppose $\sigma_1 :G\rightarrow V_1$ and $\sigma_2 :G\rightarrow V_2$ are two linear representations of $G$ on $V_1$ and $V_2$, then $\sigma_1\otimes\sigma_2 :G\rightarrow V_1\otimes V_2$ is a representation of $G$ on $V_1\otimes V_2$.  (The tensor product $W^1_i\otimes W^2_j$ of two irreps $W^1_i$ and $W^2_j$ of $V_1$ and $V_2$, respectively, need not be an irrep of $V_1\otimes V_2$, however.)

Now consider the unitary operator $\hat{U}$ on a finite dimensional vector space $V$ which has a {\it group of symmetries} $G$. This means that the matrices $\sigma(g)$ representating the elements of $g\in G$ on $V$ all commute with $\hat{U}$:  $[\sigma(g),\hat{U}]=0$.  Since $\hat{U}$ is unitary, we can decompose $V$ into a direct sum of eigenspaces of $\hat{U}$:
\begin{equation}\label{U_eig}
V=U_1\oplus U_2\oplus\cdots\oplus U_m .
\end{equation}
We can also decompose $V$ into a direct sum of irreps of $G$:
\begin{equation}
V=W_1\oplus W_2\oplus\cdots\oplus W_k .
\end{equation}
It can be shown using Schur's lemma (see \cite{Sternberg}) that since $G$ is the group of symmetries of $\hat{U}$, each irrep of $G$ must lie entirely inside some eigenspace of $\hat{U}$. Therefore, for some $i$ and $j$, if $W_i\subset U_j$ then the degeneracy of $U_j$ is at least equal to the dimension of $W_i$. Using this fact, we show now that if a graph has sufficient symmetry (in a particular sense), then it will lead to quantum walks with infinite hitting times.

\subsection{Discrete time walks on Cayley graphs}

It was observed in Sec.~II that the direction-preserving automorphism group of a Cayley graph is the group of left translations of the group elements. Since every group element corresponds to a vertex of a Cayley graph and every vertex corresponds to a basis element on the vertex Hilbert space ($\mathcal{H}^v$) of the walk, the automorphism group of a Cayley graph is a regular representation of $G$ on $\mathcal{H}^v$. Every direction-preserving automorphism of a Cayley graph will induce a representation on this Hilbert space which looks like $\sigma(g)\otimes \hat{I}_c$, where $\sigma(g)$ is the regular representation of $G$ on $\mathcal{H}^v$ and $\hat{I}_c$ is the identity on the coin space. (Note that $\sigma(g) \otimes \hat{I}_c$ is {\it not} a regular representation of $G$.) In order to prove that this walk has an infinite hitting time for certain initial states, we need to show that $\hat{U}$ has at least one degenerate eigenspace whose dimension is greater than the dimension of the coin. Since every irrep lies completely inside an eigenspace, if one of the irreps occurring in $\sigma\otimes\hat{I}$ has a dimension greater than the dimension of the coin, then we can say that the eigenspace containing that irrep has a degeneracy greater than the dimension of the coin. We now show that every irrep of $G$ occurs in $\sigma\otimes\hat{I}$. We have,
\begin{equation}
\tr(\sigma\otimes\hat{I}) = d\chi ,
\end{equation}
where $\chi=\tr(\sigma)$ and $d$ is the dimension of the coin. If $\chi_i$ is any irreducible character of $G$, then
\begin{equation}
(\tr(\sigma\otimes\hat{I})|\chi_i) = d(\chi|\chi_i) .
\end{equation}
Since $\chi$ is the character of the regular representation of $G$, $(\chi|\chi_i)\neq 0$ for any irreducible character $\chi_i$. Therefore, all irreps of $G$ occur in $\sigma\otimes\hat{I}$, and if any irrep of $G$ has a dimension greater than the dimension of the coin then there is an eigenspace of $\hat{U}$ whose dimension is greater than the dimension of the coin.  So for Cayley graphs, a sufficient condition for any discrete-time quantum walk to have infinite hitting times for some initial conditions is that the group used to define the graph have an irrep with dimension greater than the degree of the graph.

For a regular graph which is not a Cayley graph, this can be modified as follows. A discrete time walk defined on a graph will have an infinite hitting time for certain initial states if at least one irrep occurring in the induced representation (on the Hilbert space of the walk) of the direction-preserving automorphism group of the graph has a dimension greater than the degree of the final vertex.  This is a somewhat more difficult to evaluate, since unlike a Cayley graph, the induced representation of the symmetry group for a general graph is not guaranteed to include every irrep.  But in principle it is not difficult to check.

For an example of a graph with infinite hitting times, consider the Cayley graph on the symmetric group $S_n$:  $\Gamma (S_n,X)$, where $X$ is a generating set for $S_n$. In order to use the form $\hat{U}=\hat{S}(\hat{I}\otimes\hat{C})$, the graph needs to be $|X|$-colored and so we chose a generating set whose elements $x$ are such that $x^2=e$, where $e$ is the identity element. For $S_n$, such a generating set is the set of transpositions.  These form the basis for the coin space, and so the dimension of the coin is the cardinality of the generating set:  $|X|=C^n_2$, where $C^n_2 = n(n-1)/2$ is the $(n,2)$ binomial coefficient. Therefore, a symmetric group $S_n$ which has an irrep with a dimension greater than $C^n_2$ will have infinite hitting times for some initial conditions for any coin matrix. It turns out that for $n\geq 6$ any symmetric group $S_n$ possesses this property \cite{Serre}, and so the corresponding Cayley graph $\Gamma (S_n,X)$ will have infinite hitting times.

All this indicates that it is not so much the {\it size} of the symmetry group that matters for infinite hitting times, but rather the {\it kind} of group, or more precisely the size and number of irreps of the group occurring in the induced representation. Consider the hypercube, which has the group $H\cong \mathcal{Z}_2^n$ as its symmetry group (if the coin has no permutation symmetry like the DFT coin). This group is Abelian and hence has only one-dimensional irreps. So one would expect the unitary evolution operator having this as its symmetry group to have very little or no degeneracy. But when the DFT coin is used the evolution operator has sufficient degeneracy to have infinite hitting times. This is because it is the degeneracy of the DFT coin, rather than the symmetry group, that makes the evolution operator degenerate. This is supported by numerical evidence which shows that when a randomly generated non-degenerate or slightly degenerate unitary coin is used instead of the DFT, the unitary evolution operator has a very small or no degeneracy. The degeneracy of the evolution operator with an Abelian symmetry group seems to come only from the coin.

\subsection{Continuous time walks}

In the prior discussion of infinite hitting times we have used the definition (\ref{ht1}) for the hitting time, which is only well-defined for discrete time quantum walks.  We have not described a suitable measurement process to define the hitting time for a continuous time walk; nor is it obvious how to do so in the quantum case, where the presence or absence of measurements has a profound effect on the dynamics.  Any notion of hitting time for the continuous case, however, must include a measurement performed on the final vertex at some time which will verify if the particle has arrived there or not.  This leaves an ambiguity in the definition of hitting time for {\it finite} hitting times, but the notion of an {\it infinite} hitting time still has an intuitive definition:  a continuous time quantum walk has infinite hitting time if, for any set of measurements on the final vertex at any sequence of times, there is always a bounded, nonzero probability that the particle will {\it never} be found at the final vertex.

Continuous time quantum walks do not have a coin matrix, and their evolution operator for undirected graphs is $\hat{U}(t) = \exp(i\hat{H}t)$ where $\hat{H}$ is the adjacency matrix of the graph. Since there is no coin, the degree of freedom at any given vertex is one dimensional. Any eigenspace of $\hat{U}$ with a degeneracy greater than one can therefore contribute to the projector $\hat{P}$ having zero overlap with the final vertex and commuting with $\hat{U}$.   Whenever a measurement on the final vertex is performed, the measurement operators will commute with $\hat{P}$ since $\hat{P}|x_f\ra\la x_f|=0$, where $|x_f\ra$ is the final vertex state.  A nonzero $\hat{P}$ necessarily means an infinite hitting time for initial states that overlap with it. Therefore, only if $\hat{U}$ is completely non-degenerate, and none of its eigenvectors have a zero overlap with the final state, will there be finite hitting times for all initial states.  This is a lesser degree of degeneracy than is needed in the discrete time case, and we therefore expect infinite hitting times to be even more common in continuous time walks than discrete time walks.

Making the connection to symmetry again, for a continuous walk to have infinite hitting times, a sufficient condition is that at least one irrep with dimension greater than one occurs in the induced representation of the automorphism group of the graph. Consider once more the example of Cayley graphs.  As discussed in the case of discrete time quantum walks above,  the induced representation of the automorphism group $G$ on the Hilbert space $\mathcal{H}^v$ is the regular representation. Since there is no coin, $\mathcal{H}^v$ is the Hilbert space of the walk. All the irreps of the group appear in this representation because it is regular. Therefore, if any of the irreps of $G$ has a dimension greater than one, then the walk will have infinite hitting times for certain starting states. Since only Abelian groups have all their irreps of dimension one, any Cayley graph defined on a non-Abelian group will have infinite hitting times for the continuous walk.

\section{Discussion}

The role of symmetry is not restricted to producing infinite hitting times. Symmetry can also be related to the exponentially {\it fast} hitting times observed in \cite{kempe,KroviBrun}. For one of the cases examined in \cite{KroviBrun}---the discrete walk on the hypercube with the Grover coin---we observe an exponentially smaller hitting time than the classical walk on the same graph. But this happens only for the symmetric initial state $|\Psi\ra=|00\cdots 0\ra\otimes\frac{1}{\sqrt{d}}\sum_i|i\ra$. Other superpositions of coin states do not have this speed-up, but rather lead to infinite hitting times, because their overlap with $\hat{P}$ is nonzero.

In Sec.~II we noted that the group of symmetries of the $n$-dimensional hypercube is $G=H\cdot S_n$ when one takes into account the direction labels, where $H$ is the normal subgroup of direction-preserving automorphisms, and $S_n$ is the permutation group on $n$ elements (in this case, the different graph directions), and is also a subgroup of $G$. It was observed in \cite{shenvi} that $|\Psi\ra$ is the simultaneous eigenstate of eigenvalue $1$ of the subgroup $S_n$ (more precisely, the simultaneous eigenstate of the representation operators of the subgroup). Since every element of $S_n$ commutes with $\hat{U}$ for the walk with the Grover coin, a state that begins in an eigenspace of the permutation group will remain in the same eigenspace at all times.  That is,
\begin{equation}
U^t|\Psi\ra = U^t \sigma(g) |\Psi\ra = \sigma(g) U^t |\Psi\ra ,
\end{equation}
where $\sigma$ is the representation of $G$ on the Hilbert space of $\hat{U}$ and $g\in S_n$. Thus, $\hat{U}^t|\Psi\ra$ is an eigenstate of $\sigma(g)$ with eigenvalue $1$.  This eigenspace has dimension $2n$.  It turns out that the final vertex with an equal superposition of coin states $|11\cdots 1\ra\otimes\frac{1}{\sqrt{d}}\sum_i|i\ra$ also lies in this eigenspace. Since the walk never escapes this subspace to explore other parts of Hilbert space, it leads to an exponentially fast hitting time.  For a measured walk, if the symmetry subgroup commutes with the measurement operators (as is true in this case), then the same argument holds. This shows that those symmetries of the graph which are passed on to the evolution operator can create subspaces to which the walk may be confined. If the final vertex has no overlap with such a subspace for any coin state, then a walk starting in that subspace will have an infinite hitting time.  Otherwise, the hitting time will be finite, or even exponentially small depending on the dimension of this subspace relative to the full Hilbert space.

Another way of seeing these exponentially fast hitting times is as an interference effect.  Due to the symmetry of the graph, many possible classical paths through the graph arrive at the final vertex with the same phase, leading to constructive interference and an enhanced probability to be found in the final vertex.  Paths that lead to undesired states, by contrast, interfere destructively.  Looked at from this point of view, infinite hitting times represent the converse effect:  for some initial states, many paths through the graph arrive at the final vertex (or its neighbors) with phases which exactly cancel out.  In this case, the particle will never be found in the final vertex, and the hitting time is infinite.

We should also point out that the conditions derived above for infinite hitting times are sufficient for a particular graph to have infinite hitting times; but they are not necessary conditions.  For example, the symmetry group of the hypercube is Abelian, and hence does not imply that the evolution operator $\hat{U}$ must be degenerate.  Nevertheless, infinite hitting times are observed for quantum walks on the hypercube, due to the fact that the choices of coin flip operator (the Grover coin or the DFT) both have their own symmetries, which increases the total degeneracy of the evolution operator.  Infinite hitting times are therefore likely to be even more common than the conditions derived here would suggest.



One can make very plausible intuitive arguments that both infinite hitting times and exponentially fast hitting times are related to symmetry.  This makes it seem likely that the ideal problem to be solved by a quantum walk would be a problem with {\it global} symmetry, but in which this symmetry is not apparent at the local scale.  While we have precise conditions for {\it infinite} hitting times, we have not yet found criteria for the existence of exponentially {\it faster} hitting times.  One possible approach would be to identify subspaces which are preserved by the evolution operator, because they are eigenspaces of graph symmetries, and which include localized initial conditions.  But it is not clear how to find this for a general graph. It is our hope that the analysis used in this paper may provide hints as to the right way to approach this problem.

\end{document}